# Monte Carlo simulations for phonon transport in silicon nanomaterials

Dhritiman Chakraborty[*], Samuel Foster, and Neophytos Neophytou

School of Engineering, University of Warwick, Coventry, CV4 7AL, UK

[*] D.Chakraborty@warwick.ac.uk

**Abstract**

In nanostructures phonon transport behaviour is distinctly different to transport in bulk materials such that materials with ultra-low thermal conductivities and enhanced thermoelectric performance can be realized. Low thermal conductivities have been achieved in nanocrystalline materials that include hierarchical sizes of inclusions and pores. Nanoporous structures present a promising set of material properties and structures which allow for ultra-low thermal conductivity - even below the amorphous limit. In this paper we outline a semi-classical Monte Carlo code for the study of phonon transport and present an investigation of the thermal conductivity in nanoporous and nanocrystalline silicon. Different disordered geometry configurations are incorporated to investigate the effects of pores and grain boundaries on the phonon flux and the thermal conductivity, including the effects of boundary roughness, pore position and pore diameter. At constant porosity, thermal conductivity reduction is maximized by having a large number of smaller diameter pores as compared to a small number of larger diameter pores. Furthermore, we show that porosity has a greater impact on thermal conductivity than the degree of boundary roughness. Our simulator is validated across multiple simulation and experimental works for both pristine silicon channels and nanoporous structures.

*Keywords:* Thermoelectrics; nanotechnology; silicon nanomaterial; porous silicon material; nanocrystalline silicon material.

## 1. Introduction

In nanostructures with feature sizes that could vary from a few up to hundreds of nanometers, phonon transport behaviour is distinctly different than in bulk materials. As a result, materials with ultra-low thermal conductivities and enhanced thermoelectric performance can be realised. More specifically, some of the lower thermal conductivities in nanocrystalline materials have been achieved in materials that include hierarchical sizes of inclusions, at the atomic size, the nanoscale, and mesoscale, thereby achieving scattering of phonons of various wavelengths and reducing phonon transport throughout the spectrum [1, 2]. Another set of promising low thermal conductivity materials are nanoporous structures based on Silicon or SiGe [2- 5]. The structure and porosity of these materials allows for ultra-low thermal conductivity- even below the amorphous limit. In order to understand the behaviour of the thermal conductivity and assist the design of such materials for thermoelectric applications, advanced simulations that can capture all geometrical features need to be employed.

In this work, we describe the development and validation of a Monte Carlo phonon transport simulator to model thermal transport in nanoporous materials. Our simulator is based on the single phonon Monte Carlo (MC) approach to solve the Boltzmann transport equation for phonons, which provides high computational efficiency, as well as accuracy [6]. Geometry induced scattering of phonons on grain boundaries, surfaces, and voids as in realistic nanocomposite materials (shown in Fig. 3 further below), which all contribute to reducing thermal conductivity, are examined; while always keeping in mind that a significant degree of crystallinity should be allowed to keep the electronic conductivity high. Thus, this work would be useful in efforts to understand thermal conductivity in disordered materials, which are being heavily studied at the moment due to their potential to be efficient thermoelectric materials [2, 5, 7].

**Nomenclature**

| | |
|---|---|
| $\varphi$ | Porosity of porous silicon material ( Area of pores/ Total simulation area) |
| $p$ | Specularity parameter |
| $\omega$ | Dispersion relation |
| $v_s$ | Velocity of sound in silicon material |
| $v_g$ | Group velocity of phonons in silicon material |
| $\Delta_{rms}$ | Roughness of the boundaries |
| $\theta$ | Angle of phonon propagation |
| $T$ | Temperature |
| $\lambda$ | Average phonon wavelength |
| $\kappa$ | Thermal conductivity |
| $L_x$ | Length of simulation domain |
| $D$ | Diameter of pores in porous silicon material |
| $<d>$ | Grain dimension |

## 2. Approach

A Monte Carlo approach has been adopted for a semi-classical particle based description of phonon transport in a 2D simulation domain. The roughness of the domain boundaries is simulated by considering a specularity parameter, $p$, which determines the change in the direction of motion and the momentum of each phonon upon boundary scattering. [8] This is also applied to the various other boundaries in the simulations - such as boundaries of voids and grain boundaries. In the case of fully specular boundaries ($p = 1$) the phonons reflect such that the angle of incidence is the same as the angle of reflection. In the case of fully diffusive boundaries ($p = 0$) the phonons reflect such that the angle of reflection is random.

Phonons enter from either direction of the simulation domain, the time it takes to traverse the entirety of the simulation domain to exit from the other side is saved as its Time-Of-Flight (TOF). In order to initialize the temperature of the domain and to allow heat flow through it during the simulations, the simulation domain is divided into 'cells' or sub-regions. Boundary cells at the left and right of the channel are given $T_H$ and $T_C$ respectively. The rest of the domain is set at $T_{BODY}$– the average temperature between of $T_H$ and $T_C$.

The phonons are initialized as per a 'single-phonon Monte Carlo' approach which differs from the multi-phonon Monte Carlo approach described in various works in the literature [8-11]. In a multi-phonon approach, a number of phonons are initialized simultaneously. Phonon paths, energy and temperature of all cells are traced simultaneously at every time step, and often periodic boundary conditions are employed to remove the effect of the limited simulation domain. In the single phonon approach one phonon is simulated at a time from the domain boundary/edge and propagates through the simulation domain until it exits at either edge. Once the phonon exits the next phonon is then initialized.

The polarization probability, frequency, velocity, and energy of each phonon is drawn from the dispersion relation $\omega(k)$, modified by the Bose Einstein (BE) distribution at the given temperature. The fit for the dispersion relation $\omega$ and the group velocity $v_g$ is obtained as by Pop et al. in [11] (shown in Fig. 1) using equations 1 and 2 below.

$$\omega(k) = v_s + ck^2 \quad (1)$$

$$\boldsymbol{v}_g = \frac{d\omega}{dk} \quad (2)$$

Repeated iterations and scattering events give a steady state thermal gradient. By calculating the net number of phonons entering or exiting the simulation domain from either side, an average phonon flux, is determined, which is used to extract the thermal conductivity. Thus, we do not extract thermal conductivity by computing the flux that crosses a particular cross section of the domain (this would require periodic boundary conditions), but by computing the net flux through the entire simulation domain. The finite size of the simulation domain is overpassed by using the average mean-free-path to scale the simulated thermal conductivity to the actual thermal conductivity for an infinite channel length. This method makes much easier the book-keeping of all phonon attributes and could be more efficient in terms of memory requirements with the same accuracy as the more 'conventionally' used methods. We note that this approach is common in electronic transport Monte Carlo simulations, and we adopt it here for phonon transport [12,13]

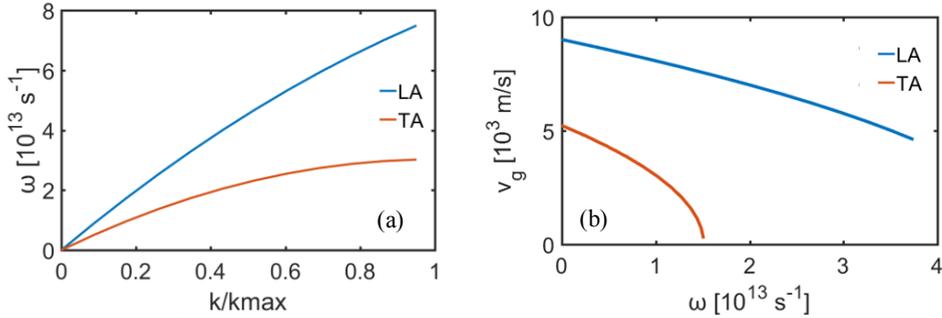

Figure 1. (a) The fit for the dispersion relation $\omega$ (equation 1) and (b) the group velocity $v_g$ (equation 2) are obtained as by Pop et al. in [11] for longitudinal acoustic (LA, blue lines) and transverse acoustic waves (TA, red lines). These are used for the initialization of random phonon group velocity, wave vector and other properties in the single phonon Monte Carlo scheme.

Phonons either scatter or are in free flight. During free flight the position $r$ at time $t$ of the phonon is given by the equation:

$$r(t_i) = r(t_{i-1}) + v_g \Delta t \quad (3)$$

Scattering of phonons is caused either by interaction with geometrical features or by three-phonon internal scattering (Umklapp processes). The three-phonon scattering, which is chiefly responsible for the change in temperature of the domain, is simulated in the relaxation time approximation and is a function of temperature and frequency, as:

$$\tau_{TA,U}^{-1} = \begin{cases} 0 & \text{for } \omega < \omega_{1/2} \\ \dfrac{B_U^{TA} \omega^2}{\sinh\left(\dfrac{\hbar\omega}{2\pi k_B T}\right)} & \text{for } \omega > \omega_{1/2} \end{cases} \quad (4)$$

where $\omega$ is the frequency, $T$ the temperature, $B_U^{TA} = 5.5 \times 10^{-18}$ s, $\omega_{1/2}$ is the frequency corresponding to $k = k_{max}/2$. These equations used here are well-established and often used to describe relaxation time in phonon Monte Carlo simulations. [6, 8-11,14]. Three phonon scattering causes a change in the energy, and thus the temperature ($T$) of the 'cell' of interaction. Every time this happens the 'cell' temperature either rises or falls. This is given by:

$$E = \frac{V}{W} \sum_{pol} \sum_{i} \left( \frac{\hbar \omega_i}{\exp\left(\frac{\hbar \omega_i}{k_B T}\right) - 1} \right) g_p DOS \Delta \omega \tag{5}$$

where, $\omega$ is the frequency, $T$ the temperature, DOS the density of states and $g_p$ the polarization branch degeneracy. Thus, a temperature gradient is established for a continuous flow of phonons (shown in Fig 2). A scaling factor (W = $4\times10^5$) is also introduced to scale the number of phonons simulated to the real population of $6\times10^{28}$ phonons m$^{-3}$ that are present at 300 K [6]. This is kept constant in our simulation domain, where the average temperature is 300 K [14].

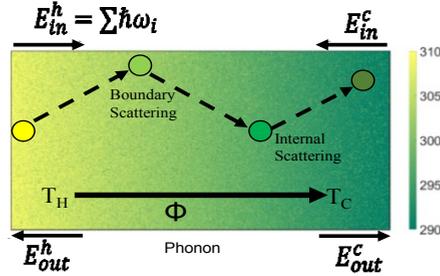

Figure 2. Simulation scheme for one phonon depicting a phonon initialized from the hot junction (T$_H$) propagating through the simulation domain undergoing internal (phonon-phonon) and boundary scattering and exiting the simulation domain to the right. The simulation domain here has the hot junction with a temperature set at 310 K (yellow) and a cold junction with a temperature of 290 K (green). 5 million phonons are inserted at both junctions and allowed to propagate through the simulation domain, and a thermal gradient is thus established. Comparing the energies of the phonons entering and exiting the simulation domain a phonon flux is obtained.

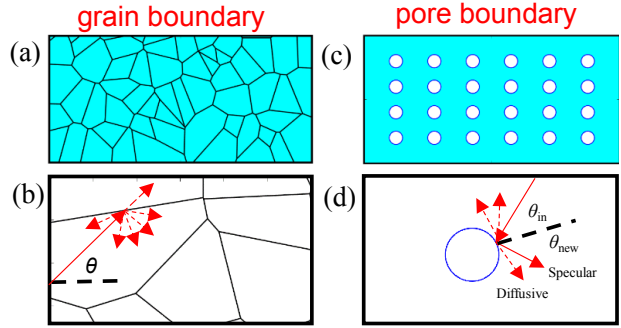

Figure 3. (a) Example of a typical nanocrystalline geometry simulated with average grain dimension <d> of 140 nm. (b) Schematic of the scattering mechanism for grain boundary scattering, indicating the initial angle of the phonon $\theta$ from the normal (black dashed line), grain boundaries (black lines), initial path of the phonon (red line) and probable paths of the phonon after scattering (red dashed lines). The probability of scattering is a function of the phonon wave vector, the roughness of the boundary $\Delta_{rms}$ and the angle of the phonon with the grain boundary $\theta_{GB}$. (c) Example of a typical nanoporous geometry simulated, for an ordered rectangular case, with porosity of about 10%. (d) Schematic of scattering mechanism for pore scattering, indicating the pore boundary (blue line), the initial angle of the phonon $\theta_{in}$, and potential new angle of propagation $\theta_{new}$ depending on specularity parameter $p$. Probable paths of the phonon after scattering for both diffusive (red dashed lines) and specular (red solid line) are also depicted. In the case of fully specular scattering, $p = 1$ and the phonon is reflected with a $\theta_{new}$ equal in magnitude to the initial angle $\theta_{in}$. In the case of fully diffusive scattering $p = 0$ and $\theta_{new}$ is random. The specularity $p$ determines the percentage specularity/diffusivity of the boundary and is a function of the phonon wave vector, and the roughness of the boundary $\Delta_{rms}$.

We consider nanocrystalline geometries as shown above in Fig. 3, where the average grain size $\langle d \rangle$ in the simulation domain defined as:

$$\langle d \rangle = \frac{L_x}{\langle N_{Gx} \rangle} \tag{6}$$

where $L_x$ is the length of the domain in the transport direction and $N_{Gx}$ is the average number of grains encountered in length $L_x$. Grains in the nanocrystalline case are generated using the voronoi tessellations function in MATLAB which creates grain boundaries by expanding grain areas radially outward from their initial "seeding points" until two areas meet [15], given initial input values for number of "seeding points" and dimensions of the domain. In these structures, thermal conductivity is impeded in two ways - the scattering of phonons due to the boundary and its roughness and internal three-phonon Umklapp scattering inside the grains. The scattering probability at grain boundaries depends on the phonon wave vector, the roughness of the boundary $\Delta_{rms}$ and the angle of the phonon with the grain boundary $\theta_{GB}$. This is given by the relation [15]:

$$p_{\text{scatter}} = \exp(-4q^2 \Delta_{rms}^2 \sin^2 \theta_{GB}) \tag{7}$$

This expression allows us to directly connect the probability of scattering to the roughness $\Delta_{rms}$ and phonon wave vector.

As briefly discussed earlier, boundary scattering in this simulation may be specular, diffusive or partially diffusive as shown in Fig. 3d. The partially diffusive condition is given by introducing a specularity parameter $p$ such that:

$$Specularity = \frac{(1-p)}{(1+p)} \tag{8}$$

where $p = 0$ gives us a completely diffusive condition and $p = 1$ a completely specular condition. The scattering mechanism at pore boundaries may also be specular or diffusive, as above. Furthermore, while the angle for the diffusive pore boundaries is random, (as shown in Fig. 3c, 3d) the angle of scattering for specular pore boundary conditions is as follows:

$$\theta_i = 2\gamma - \theta_{i-1} \tag{9}$$

where $\gamma$ is the angle between the perpendicular at the point of interaction and the x-axis of the domain, and $\theta$ is the angle of propagation of the phonon. Phonons are injected from both ends of the simulation domain, at their respective junction temperatures, to establish a temperature gradient across the device. An average of 5 million phonons were simulated in the domain to establish a convergent thermal gradient (Fig. 2).

The temperature differences used to establish the gradients were kept small and as a function of the body temperature, $\Delta T = 0.05 T_{BODY}$ [6]. This is since the relaxation time approximation used in our approach only holds for small deviations from the equilibrium temperature - as long as $\Delta T$ is kept small compared to average body temperature or $T_{BODY}$ [11]. Keeping a fixed, larger $\Delta T$ yields acceptable results at higher values of $T_{BODY}$ (say 300 K), but at low temperatures this becomes comparable to $T_{BODY}$. In that case, our simulations show that the thermal conductivity can be significantly underestimated if the same $\Delta T$ is used. (by roughly a third to half for a larger $\Delta T$ of 20 K at temperatures under 100 K). Keeping $\Delta T$ always significantly smaller than $T_{BODY}$ the characteristic peak for thermal conductivity of silicon at low temperatures is also readily obtained [6, 14].

Next, since the length of the simulated domain ($L_x$) was smaller than some phonon wavelengths, especially at lower temperatures, a scaling of the simulated thermal conductivity ($\kappa_s$) is needed to compute the final thermal conductivity $\kappa$ as [16]:

$$\kappa = \kappa_s \frac{(L_X + \lambda_{avg})}{(L_X)} \tag{10}$$

where the average phonon mean free path of Si. At room temperature, for example, $\lambda_{avg}$ = 135 nm [16]. Thus, the finite size of the simulation domain is overpassed by using the average mean-free-path to scale the simulated thermal conductivity to the actual thermal conductivity of an infinite channel length. This scaling is important for the pristine bulk case of silicon where a large number of phonons have mean-free-paths larger than the simulation domain size [14]. This is especially the case in the low temperature range where the low temperature peak of silicon is observed only after this scaling. It allows us to simulate shorter channels, and even avoid the use of periodic boundary conditions, which simplifies the simulation considerably. In the case of nanocrystalline and nanoporous structures, on the other hand, where the scattering length is determined by the grain sizes and pore distances, this scaling is not particularly important. Note that for computational ease we only use an average $\lambda_{avg}$ value, although the mean-free-path is wavevector dependent. Separate contribution of each wavevector in the thermal conductivity in a domain where phonons undergo three-phonon processes and interchange wavevectors cannot trivially be determined.

## 3. Validation and comparison to literature

All validation of our simulator has been carried out using a fixed simulation domain of length 1000 nm and width 500 nm. Initially simulations were carried out to compare and validate the simulator for bulk values of silicon thermal conductivity. The characteristic thermal conductivity peak at low temperature is obtained after scaling the simulation results using equation 10. In Figs. 4, 5 and 6 we present a comprehensive data collection of experimental (in red lines) and theoretical (in green lines) values for the thermal conductivity of Si and silicon nanostructures. Figure 4 shows that there is good agreement between our simulated results (blue line) and literature values of silicon thermal conductivity across a large temperature range with several works in the literature.

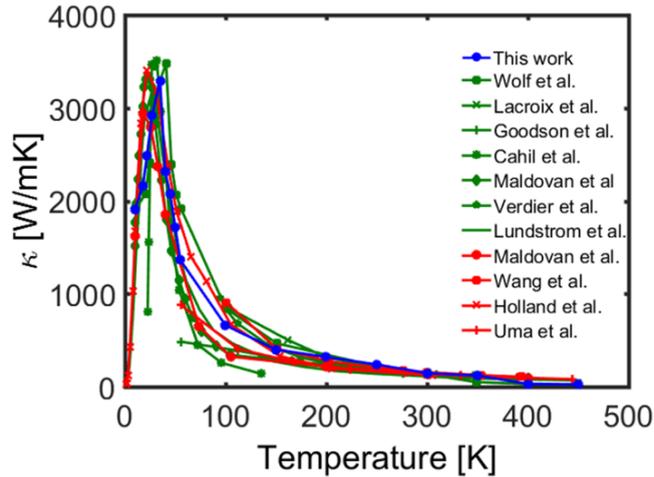

Figure 4. Validation of simulator for thermal conductivity ($\kappa$) values for the bulk case of silicon. Comparison with various simulation results (green lines) [6, 10, 15-19], and experimental results (red lines) [21-23] for silicon, available in the literature, are shown. The blue line shows our simulation results which gives the values of thermal conductivity for silicon over a temperature range of 10 K to 450 K. The simulated results are in close agreement to other literature results.

Validation was then carried out for nanocrystalline and nanoporous silicon materials (Fig. 5 and Fig. 6, respectively). Due to the large variety of simulation data for different structures that we identified in the literature, the results vary significantly, however, our results fall within the range of existing literature data, and qualitatively show the same trends. In Fig. 5 our simulations (blue line) give the values of thermal conductivity for nanocrystalline silicon structures with grain sizes <*d*> from 1000 nm down to 50 nm, for a boundary roughness $\Delta_{rms}$ = 1nm. Here, in our simulations the scattering at the grain boundaries is treated by a scattering probability $p_{scatt}$ which is a function of the phonon wave vector, the roughness of the boundary $\Delta_{rms}$ and the angle of the phonon with the grain boundary $\theta_{GB}$ as given by equation 7. In the nanoporous case (Fig. 6), we vary the percentage porosity (area of pores as a percentage of total area of domain) from 0% (an empty, pristine channel) up to 50%. Results vary depending on if the pores are ordered (solid-blue lines) or disordered (dashed blue line), however, again qualitatively and quantitatively our results are within other literature data.

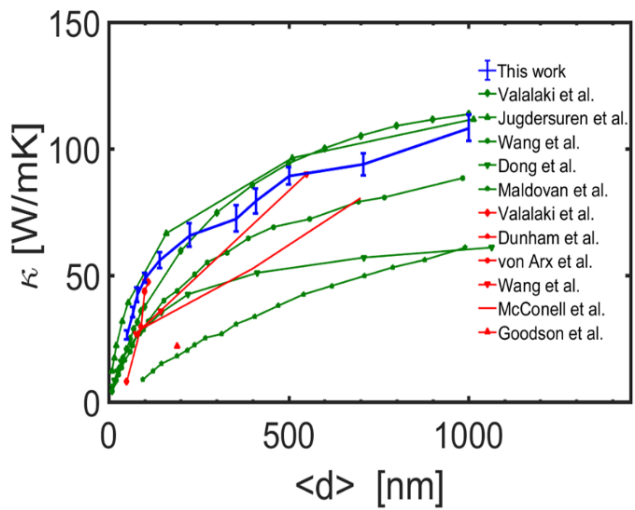

Figure 5. Validation of the Monte Carlo phonon transport simulator for nanocrystalline materials through comparisons with various simulation results (green lines) [22, 25-28], and experimental results (red lines) [17, 22, 25, 29-31] for silicon available in the literature, are shown. Our simulation results showing the thermal conductivity for silicon as the grain size is varied from a grain dimension <*d*> of 1000 nm to 50 nm are shown by the blue line.

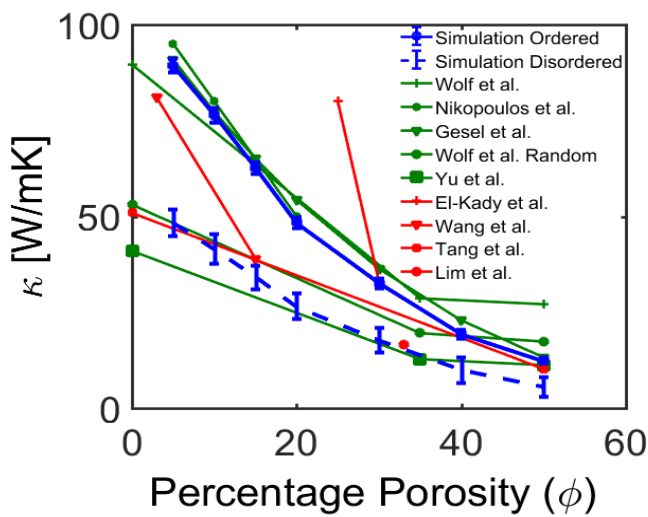

Figure 6. Validation of simulator for nanoporous material in comparison with various simulation results (green lines) [6, 14, 32-35], and experimental results (red lines) [3, 22, 34-36], for silicon available in the literature. Our simulation results for the thermal conductivity of silicon as the percentage porosity ($\varphi$) (area of pores as a percentage of total area of domain) is varied up to 50% are shown by the blue line (ordered case), and the blue-dashed line (disordered case).

## 4. Results and Discussion:

The first study we perform is to investigate the effect of the porosity ($\varphi$) of the simulation domain on the thermal conductivity by modifying the size of the pores (diameters $D$) while keeping the total number of pores in the simulation domain the same. In Fig. 7 we show that by increasing pore diameter from $D = 50$ nm to $D = 80$ nm (Fig. 7a-d, respectively) the porosity changes from ~15% to ~50% and causes a ~40% drop in $\kappa$. The specularity of the boundaries is also varied to study its influence on the thermal conductivity. Three different values of specularity: $p = 1$ (blue line), $p = 0.5$ (green line) and $p = 0.1$ (red line) are studied. An order of magnitude decrease in $p$ from $p = 1$ to $p = 0.1$ initially causes a ~30% drop in $\kappa$, but this effect diminishes as porosity increases. The lowest $\kappa$ is expected for $p = 0.1$ due to the higher diffusivity increasing randomness of phonon direction of propagation.

Next we examine the effect on $\kappa$ upon changing the number of pores but keeping the same porosity. Typical geometries for a rectangular arrangement of pores giving the same porosity for different pore sizes for $D = 60$ nm and $D = 10$ nm are depicted in Fig. 8a and 8b, respectively. Here we use a fixed specularity $p = 0.1$. It is observed that reduction in pore diameter, which leads to an increase in the number of pores at the same porosity, leads to lower thermal conductivity. This effect is observed since an increase in number of pores would lead to an increase in the surface area available for scattering – thus increasing TOF and leading to a decrease in $\kappa$ as illustrated in Fig. 8c. However, as the percentage porosity increases, the difference in $\kappa$ due to pore diameter is greatly minimized. At higher porosities pores move much closer together, reducing the space available for phonons to travel through and increased scattering is observed regardless of pore diameter. This leads to similar thermal conductivity values for pores of different diameters at high porosities.

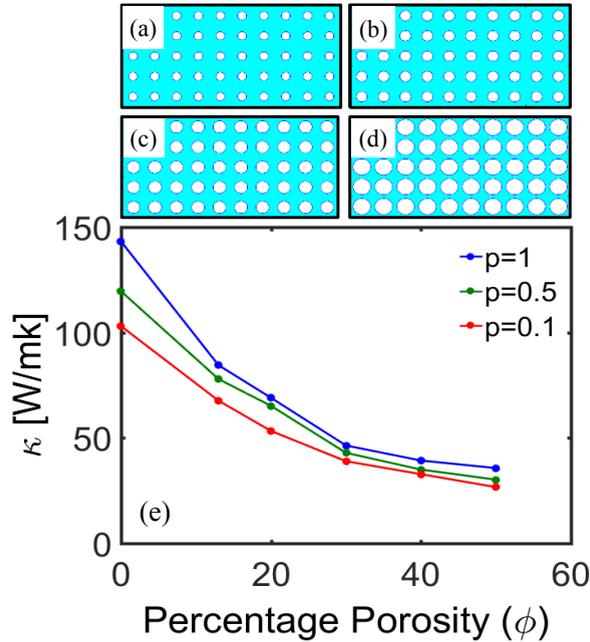

Figure 7. The effect of pore size only on the thermal conductivity ($\kappa$) is examined. Pore diameters ($D$) are varied while the pores are kept in the same positions. Increasing pore diameter from $D = 50$ nm to $D = 80$ nm, (a) to (d) respectively, changes the porosity by over 35% and causes a 40% drop in $\kappa$. This is examined for three different values of specularity $p = 1$ (blue line), $p = 0.5$ (green line) and $p = 0.1$ (red line).

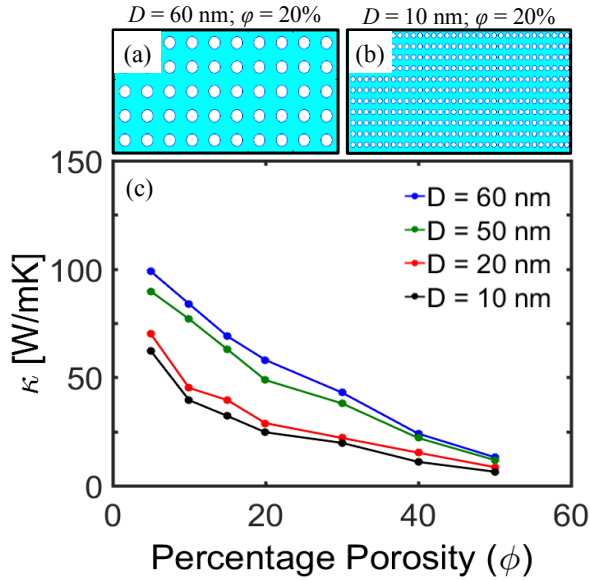

Figure 8. Effect of pore size and density on ($\kappa$) is examined. Here, a certain porosity, is achieved by utilizing pores of different sizes ($D$ = 60 nm, 50 nm, 20 nm, and 10 nm depicted by the blue, green, red, and black lines, respectively). Typical geometries for a rectangular arrangement of pores giving the same porosity for different pore sizes $D$ = 60 nm and $D$ = 10 nm are depicted in (a) and (b) respectively. These are for a fixed specularity $p$ = 0.1. It is observed that reduction in pore diameter leads to lower thermal conductivity at the same porosity. However, this effect diminishes as porosity increases.

## 5. Conclusions

In this work we described some features of a single particle Monte Carlo phonon transport method and validated a simulator we constructed based on that method to numerous experimental and simulation data. We then investigated the thermal conductivity in silicon nanostructured materials. We investigate the effects of pore and grain boundary scattering on the thermal conductivity. We show that the effect of porosity is stronger than that of boundary roughness. We also show that at the same porosity, a larger decrease in thermal conductivity is achieved by increasing pore number density than by increasing the pore diameter. An almost 50% decrease in thermal conductivity is observed with a decrease in pore diameter from $D$ = 60 nm to $D$ = 10 nm at the same porosity (observed for porosities from 15 % to 25%); however, this effect diminishes at higher porosities ($\varphi$ = 35% and above). Our results would be useful in the design of nanostructured thermoelectric materials with ultra-low thermal conductivities.

Acknowledgements: This work has received funding from the European Research Council (ERC) under the European Union's Horizon 2020 Research and Innovation Programme (Grant Agreement No. 678763).

# 6. References

[1] K. Biswas, J. He, I.D. Blum, C. Wu, T. P. Hogan, D. N. Seidman, V. P. Dravid, and M. G. Kanatzidis, Nature (2012): 489.7416: 414-18.
[2] J. A. P.-Taborda, M. M.-Rojo, J. Maiz, N. Neophytou, M. M. González, Nature Sci. Rep., (2016): 6, 32778.
[3] J. Tang, H. T. Wang, D. H.Lee, M. Fardy, Z. Huo., T.P. Russell.,P. Yang, Nano letters, (2010) 10(10), 4279-4283.
[4] J.Lee, J. Lim, P.Yang, Nano letters (2015):15.5: 3273-3279.
[5] B. Lorenzi , D. Narducci, R. Tonini, S. Frabboni, G.C. Gazzadi, G. Ottaviani, N. Neophytou, X. Zianni, Journal of Electronic Materials, (2014) 43.10: 3812-3816.
[6] S. Wolf, N. Neophytou, and H.Kosina, J. Appl. Phys. (2014): 115, 204306.
[7] N. S. Bennett, D. Byrne, A. Cowley, N. Neophytou, Appl. Phys. Lett. (2016) 109, 173905.
[8] Mazumder, Sandip, and Arunava Majumdar, Transactions-American Society of Mechanical Engineers Journal of Heat Transfer 123.4 (2001): 749-759.
[9] Pop, Eric, Sanjiv Sinha, and Kenneth E. Goodson, Proceedings of the IEEE 94.8 (2006): 1587-1601.
[10] Lacroix, David, Karl Joulain, and Denis Lemonnier Physical Review B 72.6 (2005): 064305.
[11] Pop, Eric, Robert W. Dutton, and Kenneth E. Goodson, Journal of Applied Physics 96.9 (2004): 4998-5005.
[12] Jacoboni, C., & Reggiani, L. Rev. Mod. Phys., (1983): 55(3), 645–705.
[13] Kosina, H., Nedjalkov, M., & Selberherr, S. IEEE (2000) Electron Devices, 47(10), 1898–1908.
[14] Wolf, S., Neophytou, N., Stanojevic, Z., & Kosina, H. (2014) Journal of electronic materials, 43(10), 3870-3875.
[15] Aksamijia, Z., & Knezevic, I. (2014). Physical Review B 90(3), 1–8.
[16] Jeong, Changwook, Supriyo Datta, and Mark Lundstrom. Journal of Applied Physics 111.9 (2012): 093708.
[17] Kenneth Goodson, Cahill, David G., and Arunava Majumdar, Journal of Heat Transfer 124.2 (2002): 223-241.
[18] Cahill, D. G., Ford, W. K., Goodson, K. E., Mahan, G. D., Majumdar, A., Maris, H. J., ... & Phillpot, S. R. (2003). , Journal of applied physics, *93*(2), 793-818.
[19] Verdier, M., K. Termentzidis, and D. Lacroix. Submicron Porous Materials. Springer International Publishing, 2017. 253-284.
[20] M Verdier, R Vaillon, S Merabia, D Lacroix, K Termentzidis Pan Stanford Publishing Pte. Ltd., 2017.
[21] Maldovan, Martin, Jean, V. Journal of Applied Physics 115.2 (2014): 024304.
[22] Wang, Zhaojie, et al., Nano letters 11.6 (2011): 2206-2213.
[23]  Holland, M. G. "Analysis of lattice thermal conductivity." Physical Review132.6 (1963): 2461.
[24] Uma, S., et al., International Journal of Thermophysics 22.2 (2001): 605-616.
[25] Valalaki, Katerina, Philippe Benech, and Androula Galiouna Nassiopoulou. "High Seebeck coefficient of porous silicon: study of the porosity dependence." Nanoscale research letters 11.1 (2016): 201.
[26] Dong, Huicong, Bin Wen, and Roderick Melnik. Scientific reports 4 (2014).
[27] Jugdersuren, B., et al. Physical Review B 96.1 (2017): 014206.
[28] Maldovan, Martin. "Thermal energy transport model for macro-to-nanograin polycrystalline semiconductors." Journal of Applied Physics 110.11 (2011): 114310.
[29] Dunham, Marc T., et al. Applied Physics Letters109.25 (2016): 253104.
[30] Von Arx, Martin, Oliver Paul, and Henry Baltes. Journal of Microelectromechanical systems 9.1 (2000): 136-145.
[31] McConnell, Angela D., Srinivasan Uma, and Kenneth E. Goodson. Journal of Microelectromechanical Systems 10.3 (2001): 360-369.
[32] Gesele, G., et al. Journal of Physics D: Applied Physics 30.21 (1997): 2911.
[33] Yu, Jen-Kan, et al. Nature nanotechnology 5.10 (2010): 718-721.
[34] Nikopolous et al. Journal of the American Ceramic Society 66.4 (1983): 238-241.
[35] El-Kady, Ihab, et al. Sandia National Labs, Albuquerque, NM, Report No. SAND2012-0127 (2012).
[36] Lim, Jongwoo, et al. ACS nano 10.1 (2015): 124-132.